# Thermal Shear Waves induced in Mesoscopic Liquids at Low Frequency Mechanical Deformation

Eni Kume & Laurence Noirez

Laboratoire Léon Brillouin (CEA-CNRS), Université Paris-Saclay, 91191 Gif-sur-Yvette Cedex, France

**Abstract:** We show that a viscous liquid confined between two low thermal conductive surfaces ($Al_2O_3$) emit a thermal response upon applying a low frequency (Hz) shear excitation. Hot and cold thermal waves are observed *in situ* at atmospheric pressure and room temperature, in polypropylene glycol layers of various thicknesses ranging from 100µm up to 340µm, upon applying a mechanical oscillatory shear strain. The observed thermal effects, synchronous with the mechanical excitation, are inconsistent with a homogeneous viscous flow. It indicates that mesoscopic liquids are able to (partly) convert mechanical shear energy in non-equilibrium thermodynamic states. This effect called thermo-elasticity is well known in solid materials. The observation of a thermal coupling to the mechanical shear deformation reinforces the assumption of elastically correlated liquid molecules. The amplitude of the thermoelastic waves increases linearly by increasing the shear strain amplitude up to a transition to a non-linear thermal behavior, similar in strain behavior to an elastic to plastic regime. The thermo-elastic effects are not detectable via stress measurements and thus appear as a complementary liquid dynamic characterization.

Keywords: confined liquids, shear elasticity, dynamic analysis, thermoelasticity.

## 1. Introduction

From large length scales (geological and even astrophysical scales) down to nanoscale confinement, fluids play crucial roles definitively at all length scales. But how to differentiate liquids from solids and does this difference depend on the scale at which the observation is done? It is conventionally accepted that the rapid molecular dynamics of (ordinary) liquids do not allow the propagation of low frequency shear waves, and thus that the (static) shear modulus is a solid-like characteristic. However, the properties of the confined material can be very different from that of the three-dimensional bulk phase. Recent experimental and theoretical developments based on scale dependent analysis challenge the rapid dynamics hypothesis and point to the possible existence of a mesoscopic liquid shear elasticity at low frequency (Fig.1a) [1-17].

We are interested in what happens in the low frequency region (~ 1Hz) when a viscous liquid is submitted to a mechanical shear strain at scales where recent developments have pointed out the existence of liquid shear elasticity (Fig.1a) [1-17]. The shear elastic dynamic behavior is typically observable in the low strain region by applying an oscillatory shear strain using the conventional dynamic analysis imposing a sine shape oscillatory shear strain: $\gamma(t) = \gamma_0 \cdot sin(\omega \cdot t)$ where $\omega$ is the frequency, $\gamma_0$ the imposed shear strain defined as $\gamma = \delta l/e$ with $\delta l$ being the displacement, $e$ the gap thickness of the confined fluid and $\omega$ is the frequency. For the polypropylene glycol (PPG-4000), the elastic behavior is visible in the dynamic response below $\gamma < 100\%$ at 100µm (Fig.1b). It is identifiable by a shear modulus G' higher than then viscous modulus G" in agreement with a nearly in-phase shear stress-shear strain waves illustrated in the left insert of Fig.1b. At large strain amplitude, the elastic-like regime vanishes being progressively replaced by a viscous regime (G" > G'), in agreement with the π/2-phase shifted stress wave (right insert of Fig.1b) indicating a purely viscous regime. At macroscopic scale whatever the strain amplitude, the liquid exhibits a viscous response (not illustrated here).

The mesoscopic measurements show thus that the viscoelastic response is not universal but is a function of the scale at which the fluid response is measured and of the applied strain [1, 9-12, 16-17]. The mesoscopic shear elasticity has been identifiable in both simple liquids (Van der Waals and H-bond liquids), complex fluids (polymer melts, molecular glass formers, ionic liquids) and physiological fluids [12].

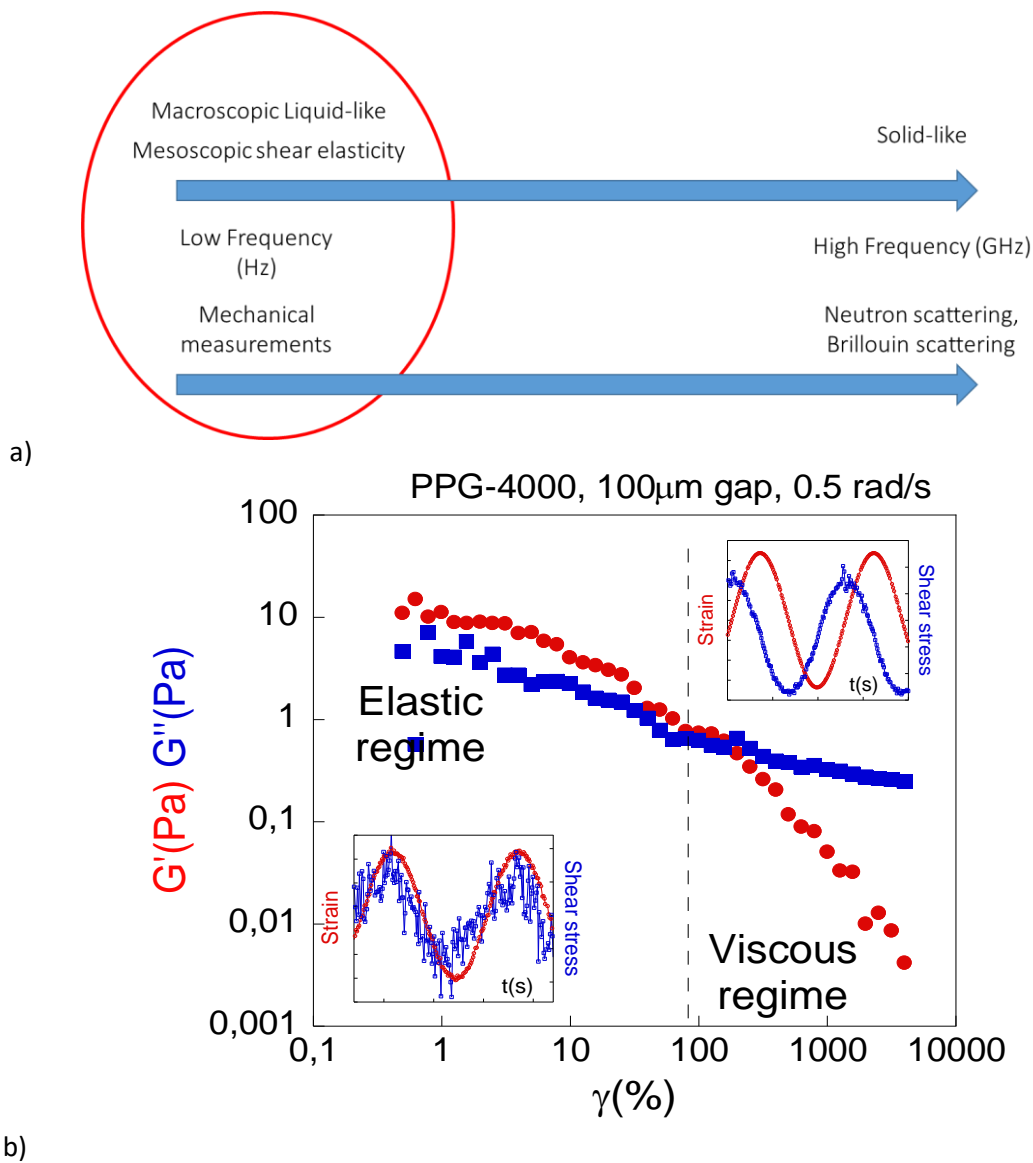

**Fig.1:** a) Scale and frequency dependence of liquid properties: We explore the dynamic properties of liquids in the low frequency domain (∼ 1Hz, represented by the red circle) where mesoscopic liquids exhibit shear elastic properties. b) Dynamic mechanical response of the polypropylene glycol (PPG-4000) in terms of shear elastic G' and viscous modulus G" at 0.5 rad/s as a function of the shear strain (γ). The elastic response is identifiable at low shear strain (G' > G") in agreement with the nearly in-phase stress- strain waves (left insert). At large strain

amplitude, the viscous response dominates (G" > G') in agreement with the π/2-phase shifted stress wave (right insert). Measurements carried out in total wetting conditions (Alumina substrate), at 100µm thickness, room temperature and far away from the glass transition (Tg = -75°C).

The shear elasticity of mesoscopic liquids is experimentally accessible amplifying the liquid interaction to the surface. High energy surfaces like the Alumina ($Al_2O_3$) lead to a complete wetting (contact angle ~ 0°)[1,2]. The strong liquid/substrate interaction amplifies the transmission of the shear strain to the fluid. In such conditions, the dynamic response exhibits at mesoscopic scale, a liquid shear elasticity in the low frequency range (0.1-10Hz) [1-7, 10-12].

Because of the shear elasticity, fluids resist to a shear field and their resistance depends on the considered scale, being reinforced when the scale decreases [1, 5, 6-7, 12]. Therefore, when a fluid is submitted to a shear strain, its elasticity might be actioned, exerting either an expansion or a compression. When an elastic material expands, its thermal energy is changed. The immediate consequence is that a thermo-elastic coupling becomes possible, challenging the assumption of an instant dissipation via thermal fluctuations and justifying the search of a thermal response.

Under the assumption of elastically correlated liquid molecules, we present a thermal approach to analyze the behavior of the liquid to a mechanical shear strain. We use the experimental conditions that have enabled the identification of the low frequency shear elasticity; i.e. low frequency oscillatory shear strain, sub-millimeter scale, total wetting substrate/fluid conditions and nearly insolating surfaces to focus the analysis on the liquid thermal behavior.

## 2. Experimental:

Recent instrumental progresses in infrared detection enable now an accurate determination of the temperature in a wavelength range of 7-14µm. We use this technique to record the *in-situ* temperature of confined viscous fluids submitted to a controlled external mechanical shear strain delivered by a strain-imposed rheometer (ARES2 - TA-Instrument) (Scheme 1).

The infra-red emissivity measurements are carried out at room temperature, in real-time conditions with a microbolometer array of 382 x 288 pixels working at 27 Hz in the range of long wave Infrared bands (LWIR), i.e. wavelengths ranging between 7 to 14 µm. The thermal emissivity is measured by radiation transfer using the Stefan-Boltzmann law: $E = e_m \cdot \sigma \cdot A \cdot (T^4 - T_c^4)$ where $E$ the radiated energy, $e_m$ the emissivity coefficient, $A$ the radiating area, $T$ the temperature of the sample and $T_c$ the temperature of the surroundings. $\sigma$ is the Stefan's constant. The microbolometer is coupled to a home-made objective to magnify the thermal image that is a combination of 12 Germanium lenses [18]. The focal of this objective is 7.5mm. The numerical aperture is F/1 and the spatial resolution ellipsoid is 0.1mm. The microbolometer array focusses the liquid confined between two surfaces. The thermal pictures are corrected from the static thermal environment by subtracting the median value measured at rest prior the dynamic measurements. Each image is processed. The images are then compacted along the Y-axis to build a kinetic picture.

The mechanical dynamic measurements were carried out using the standard procedure for a rheology experiment, consisting in filling the gap between two co-centred disks, one surface driven by an oscillatory motion with respect to the disk centre of perfectly controlled frequency ω and of amplitude γ, the other one fixed and (scheme 1). The liquid periphery is free and in equilibrium with the atmospheric pressure. The rheometer is equipped with a normal force sensor enabling the determination of the variation of the normal force with a precision of +/- 0.1g.

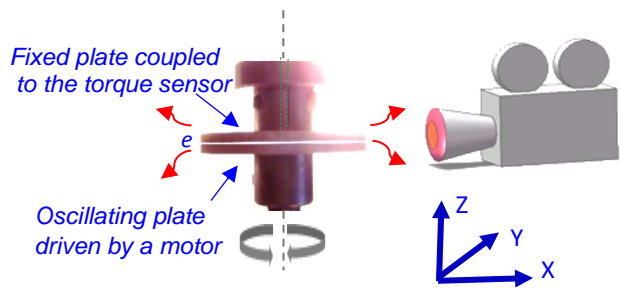

Scheme 1: Set-up: The liquid fills the gap between two coaxial discs made of Alumina: one of which is driven in an oscilllatory rotational motion and the other is fixed and coupled to a torque sensor. The two coaxial surfaces are separated here from e = 0.100mm (photo). At right, an infra-red camera equipped with a macro-lens is positioned so as to observe the liquid layer in the shear plane (xOz).

The liquid, polypropyleneglycol-4000 (Sigma-Aldrich manufacturer), has a molecular weight of 3500-4500 g/mol, a glass transition at Tg = -75°C, nearly no vapor tension (< 0.01mm Hg at 20°C). Its viscosity at room temperature and macroscopic scale is $\eta$ = 100 mPa.s. The low frequency range probed in this study rules out any coupling with molecular relaxation times [19]. The liquid is submitted to an oscillatory shear strain using the standard dynamic mechanical analysis. A sine shape oscillatory shear strain is imposed following the formalism: $\gamma(t) = \gamma_0 \cdot sin(\omega \cdot t)$ where $\omega$ is the frequency and $\gamma_0$ the imposed shear strain [8, 9]. All the measurements are carried out at room temperature, atmospheric pressure and without external heat source. The transmission of the stress from the surface to the sample is reinforced by using high energy alumina fixtures of 45 mm diameter [1,2]. The excellent wetting procured by the alumina substrate strengthens the interaction of the liquid molecules to the surface. The high affinity to the substrate reduces the interfacial gas layer trapped between the liquid and the substrate ("pancake" effect). The relative low thermal conductivity of Alumina surfaces (~30W/mK) enables to work in nearly adiabatic conditions with respect to the experimental time scales.

## 3. Results:

3.1 Revealing a thermo-mechanical coupling in mesoscopic liquids and its elastic origin:

To examine the thermal properties, the liquid is probed at low frequency (0.5-1 rad/s). The lowest shear strain providing an exploitable thermal image is 200% (microbolometer limitations). Following Fig.1b, these strain conditions correspond to the entrance in the viscous regime.

It is observed that the temperature of the liquid is no longer homogeneous during the oscillatory shear stress but that it presents local thermal variations under conditions for which a viscous behavior is expected (Fig.1 b). These thermal variations are positive and negative, localized in space and time. They divide the fluid in nearly three bands parallel to the plates called the upper and the bottom bands, and the middle band (Fig.2a).

The thermal waves vary alternatively from cold to hot along the oscillatory period. They reproduce the mechanical waveform of the shear strain input (left insert of Fig.2b) and can be modeled as: $\Delta T(t) = \Delta T_A \cdot sin(\omega \cdot t + \Delta\varphi)$ where $\Delta T_A$ is the amplitude of the thermal wave and $\Delta\varphi$ is the phase shift with respect to the shear strain wave and $\omega$ the frequency of the mechanical excitation (the median part is not illustrated, the temperature variation being below the accuracy). The thermal wave is reversible and stable over time. We see that the hot and the cold parts of the thermal wave are symmetric for different strain values, meaning that the sine waveform holds true as the deformation is increased, showing the linearity of the phenomenon. Thus, we may propose that the temperature oscillates symmetrically around a temperature which is the equilibrium temperature, meaning that there is no exchange with the environment. For a thermal variation of 0.08 K, hot and cold parts have nearly the same value of 0.04 K in the regime of linear thermal response. The thermal signal can be modeled by a sin wave of same period as the oscillatory strain.

Another remarkable result is that a thermal wave is recorded in the so called viscous regime. Fig.2b shows the evolution of the temperature (full amplitude value from maximum to minimum temperature (peak to peak value)) in arbitrary bands selected to represent the main different thermal behaviors as the function of the strain amplitude. The average temperature in the gap is also represented. The study of the phase shift between the imposed shear strain and the thermal waves is also instructive.

- At low strain amplitude ($\gamma < 1000\%$), the phase shift is negligible $\Delta\varphi < 10\%$, indicating a nearly instant response, in agreement with a "pure" elastic behavior.

- At moderate strain amplitude ($1000\% < \gamma < 2000\%$), the phase shift increases nearly linearly up to a plateau while being always less than $\pi/4$.

- At large strain amplitude ($\gamma > 2000\%$), the phase shift reaches a value around 42°. $\Delta\varphi$ does not evolve anymore meaning that the thermal response remains mainly in phase with the imposed shear strain over the entire accessible range of amplitude.

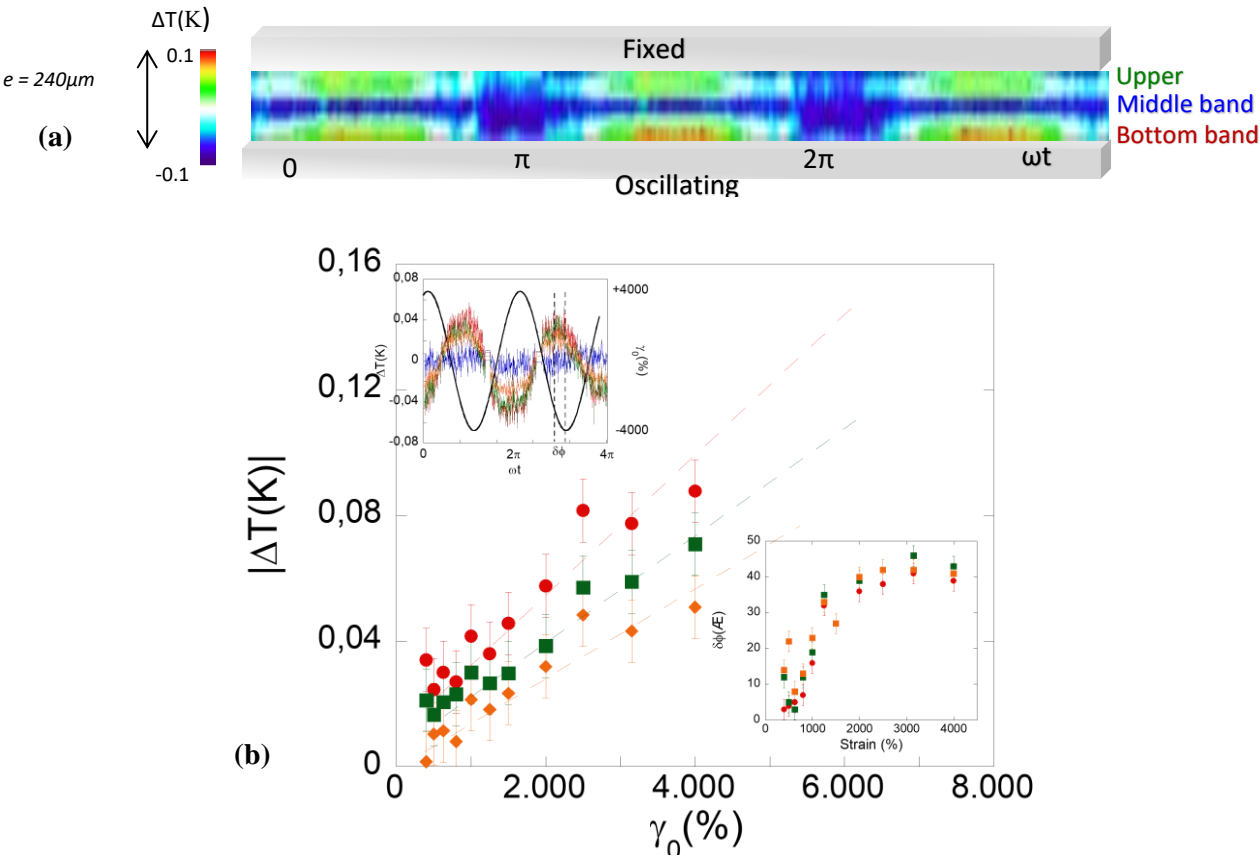

**Fig.2:** a) Thermal response of the liquid PPG-4000 recorded along three oscillatory period (ω = 0.5 rad/s, γ = 2000%, and e = 240 μm gap thickness, room temperature measurements carried out on alumina plates, the upper plane was fixed while the bottom one oscillated). b) Strain dependence of the maximum of the temperature variation amplitude (Peak-to-peak amplitude) |ΔT(K)|. Sample: PPG-4000 at gap thickness 0.240mm, ω=0.5 rad/s, as extracted from the sin harmonic fit - measurements below 400% are below the accuracy. The left insert details the thermal waves recorded at γ = 4000%, at the same gap thickness (240μm) and frequency (ω=0.5rad/s). The right insert illustrates the phase shift as a function of the shear strain. The color code is the same for the three figures:  Bottom band: (●), upper band: (■) and total gap: (◆).

Fig.2b shows that hot and cold thermal bands exhibit an almost linear dependence to the shear strain amplitude up to γ < 4000% while the phase shift never exceeds π/4 (Fig.2b). These characteristics are those indicating a thermoelastic coupling which is in agreement with a previous study [20 and references therein]. While the thermal signal is hardly accessible below γ < 200%, the linearity of the dependence of the thermal effect at larger strain amplitude is a strong indication that the thermo-elastic mechanism is occurring from the smallest values of shear strain. Let's describe the frame of a classical thermoelastic behavior.

In shear geometry, the shear strain is given by $\gamma(z) = \delta l(z)/e$ with $\delta l$, the displacement of one surface, $e$ the distance between the two surfaces (the gap) and $z$ the considered height in the gap. For simple unidimensional case, initially at the temperature $T_0$, a uniaxial stretching gives rise to $\gamma(z) = \alpha \cdot |(T-T_0)|$ where $T$ is the temperature, $T_0$ the reference temperature (here $T_0$ is the room temperature) and $\alpha$ the thermal expansion coefficient. In the linear region of the thermal response, the ratio $K_{thermo-elastic} = |(T-T_0)|/\gamma(z)$ is a constant of nearly similar absolute value for hot and cold thermal bands ($K_{thermo-elastic} \sim 0.18 \cdot 10^{-2}$K). It represents the thermal analogue of the shear stress reported to the shear strain, $\sigma/\gamma$ which defines the shear elastic modulus according Hooke's law. Therefore the thermal study might visualize local elastic responses of the liquid; i.e. its capability to change locally its pressure upon mechanical excitation (the corresponding pressure variation for a temperature variation of $10^{-2}$K is of the order of MPa). The thermal approach reveals at large strain amplitudes, an elastic-like behavior that is not identifiable via a viscoelastic dynamic analysis which indicates a viscous behavior(Fig.2b).

3.2 The scale dependence:

Since we have understood that hot and cold thermal waves exhibit similar characteristic and are both sides of the same elastic-like mechanism that is observable when the liquid is confined, we examine how this dynamic effect evolves with the sample size. The peak-to-peak value is an interesting parameter since it corresponds to the difference between its highest positive (hot) peak and its lowest negative (cold) peak. It refers to the maximum change occurring during one cycle. Fig.3 shows the evolution of the thermal response as a function of the strain at different sample thicknesses from 340μm down to 100μm at 1 rad/s.

Up to shear strain values γ < 2000%, a similar sine-like variation of the temperature is observed in each band (Fig.3 a-d). The amplitude of the thermal waves and thus the average gap temperature exhibits a linear relation with increasing strain up to around γ = 2000%. A deviation from the linearity is clearly observed at higher strain values for the smallest gaps with a pronounced lowering of the thermal variation at lower gap thickness (Fig.3d). The deviation from linear thermal response is also dependent of the applied frequency; the comparison of Fig.2b and Fig.3 shows that the transition occurs at smaller strain rates at 1 rad/s compared to 0.5 rad/s (Fig.2b) for which the instrumental limitations do not enable to probe larger strain amplitudes.

As reported in different papers [1, 6-8, 16-17], the shear elasticity is reinforced at low scale (in Figure 5 of [10], the shear elasticity of PPG reaches 200 Pa). The low thickness behavior is particularly interesting since a phenomenological analogy could be done with the strain behavior of solids under strain, showing a reversible, linear deformation in the elastic regime and, at larger strain values, a plastic-like regime of deformation observable in most solid materials (Fig.3f)[21]. In this "plastic-like" regime, the thermal modulation is distorted and exhibits harmonics [20]. However, no specific interfacial effects like interfacial surface slip is visible on the basis of the thermal analysis, thus indicating that the thermal bands are related to a bulk mechanism; a variation of the liquid density within the limit of the liquid compressibility. This mechanism is likely precursor of shear induced isotropic to nematic phase transitions in liquid crystalline fluids, those present also a scale dependence [22] and probably also related to or precursor of shear banding instabilities identified in various solutions interpreted as shear induced concentration changes only [23].

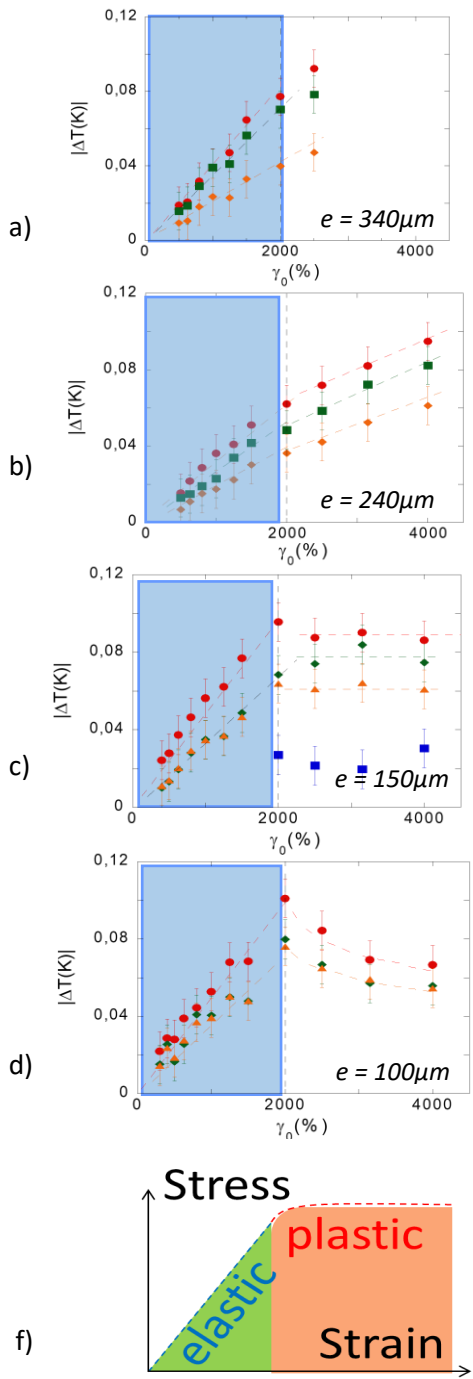

**Fig.3**: Influence of the gap thickness on the thermal response (Peak-to peak value) of the liquid upon increasing oscillatory shear strain at 1 rad/s (PPG-4000, room temperature measurements, total wetting conditions (Alumina)): Bottom band: (●), middle band: (■), upper band: (◆) and total gap: (▲). a) e= 340µm., b) e= 240µm, c) e= 150µm, d) e= 100µm, f) schematic representation of the typical solid-like behavior submitted to an external stress exhibiting elastic behavior at low strain and plastic at large one. The thermal variation of the middle band being much lower than the other bands *("neutral temperature band")*, it is not systematically represented. Large shear strain amplitudes are not experimentally accessible at 340µm.

While observed in the so called viscous regime, the thermal wave in phase with the applied strain wave exhibit elastic-like characteristics. Therefore it would be of interest to examine a secondary effect of shear strain in solids that are normal forces.

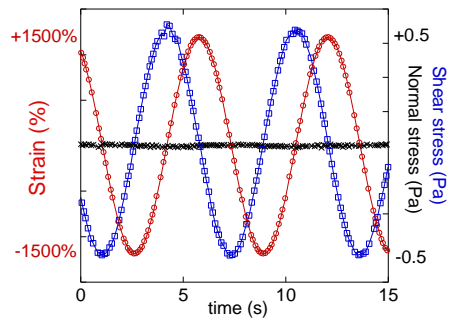

**Fig.4**: Shear stress (blue squares) and shear strain (red circles) waves and normal stress (black crosses) for PPG-4000 at 0.340mm gap thickness, at ω = 1 rad/s and γ = 1500%, (thermal waves are nearly in phase with the imposed shear strain). The shear stress is π/2 phase shifted with respect to the shear strain in agreement with a viscous regime.

For a normal force to develop, it is necessary to couple shear stress and normal force. This condition is fulfilled for solid materials; the hypothesis is that a three–dimensional deformation produces a 'strain-stiffening' effect. In the present case, normal stresses represent less than 0.3% of the shear stress and are about 6 times below the sensor accuracy given by the manufacturer (Fig.4). The normal force is however negligible with respect to the variation of the shear stress.

## 4. Conclusions:

We have shown that thermal and stress mechanical approaches are complementary ways to characterize mesoscale liquid properties. The comparison of thermal and stress data shows similarities but also fundamental differences. Shear elasticity in confined liquids has been typically measured in the low strain regime where the mechanical perturbation is the lowest. The shear elasticity is no more accessible at large strain deformation [1-4, 10]. However, the thermo-elastic signal is still observed in the so called viscous region of the strain-stress region indicating that the liquid responds elastically even in the viscous region, while the elastic component is no more observable in the stress measurement. The thermal approach is thus unique in the sense that it evidences indirectly that at large strain amplitudes, the confined liquid has not shear melt its elastic properties as it is expected in a plastic behavior, but still exhibits an elastic response via its thermal signature and while the stress measurements indicate a viscous behavior (Fig.1b).

The thermal effects detailed here reinforce the interpretation of the dynamic properties of the confined liquid in terms of elastic correlations. The ability of the liquid to convert the mechanical (shear) wave in local symmetrical hot and cold thermal waves oscillating around the equilibrium temperature while maintaining the waveform and frequency of the mechanical excitation, to exhibit at moderate shear strain a linear dependence of the thermal amplitude with the shear strain, is undoubtedly an elastic characteristic. A full and instantaneous dissipation of a mechanical action (low frequency) in the noise of thermal fluctuations [24] rules out but indicates that the thermal fluctuations are dynamically correlated.

Therefore, the viscous regime is thus more complex than usually accepted; the thermal variation indicates indeed that the liquid has an ability to store dynamically the mechanical energy associate to the shear strain. It proves that shear waves propagate in the liquid, which is a solid-like characteristic. The increase in internal energy due to the mechanical action is also correlated to a slight change of intermolecular distances. This change modifies the entropy of the system with possibly a very slight ordering (in agreement with the existence of a cooling state).

We have also shown that at very large shear strain amplitude, the thermal effect does not evolve anymore and collapses. This is particularly visible at the smallest gaps (100µm) where the shear elasticity is reinforced. The scale dependence is also found experimentally and theoretically for the shear elasticity [1, 3-8, 10, 16-17, 20].

The scale dependence of thermal and elastic properties require unconventional theoretical considerations. The liquid shear elasticity has independently established by both an experimental and theoretical approaches; indeed the scale dependence is in agreement with new theoretical models, foreseeing that liquids can support a limited propagation of shear waves well above nanoscopic scales [13-17]. Liquid elasticity can be also understood in the frame of the non-affine models developed to quantitatively predict elastic and viscoelastic constants in glasses of polymers and colloids (NALD approach [25, 13]). A $L^{-3}$ law has been experimentally verified on the basis of published data, for a wide range of fluids at the sub-millimeter scale (glycerol, ionic liquids, polymer melt, isotropic liquid crystals) [16]. It is also found to be in line with the moduli of elasticity published by the pioneering Derjaguin at the scale of several microns [6-7] or even at the nanoscopic scale probed by AFM by E. Riedo [3], therefore over a very wide dimensional range.

This novel and scale-dependent approach combining both thermal and dynamical properties is certainly a promising avenue for a better understanding of the complexity of fluids, especially in confined geometry, which typically corresponds to microfluidic conditions and the search for new small-scale dynamic properties.

**Acknowledgments:**
The authors are very grateful to Patrick Baroni to instrumental developments and innovations. They also thank warmly Alessio Zaccone for theoretical modeling of the shear elasticity. This work received funding from the European Union's Horizon 2020 research and innovation program under the Marie Sklodowska-Curie grant agreement no. 766007 and LabeX Palm (ANR-11-Idex-0003-02).

Polymers: Identification of Gap Thickness and Slipping Effects, Langmuir 25 (9),(2009) 5248-5252. DOI: 10.1021/la803848h.

23. S. Fieldings, P.D. Olmsted, Early Stage Kinetics in a Unified Model of Shear-Induced Demixing and Mechanical Shear Banding Instabilities, Phys. Rev. Lett. 90 (2003) 224501-1.

24. Hydrodynamic Fluctuations in Fluids and Fluid Mixtures, Chap. 3, Fluctuations in fluids in thermodynamic equilibrium, authors: M.Ortiz de Zárate, J. V.Sengers, Elsevier (2006), 39-62.

25. Zaccone A., Scossa-Romano E. Approximate analytical description of the nonaffine response of amorphous solids, Phys. Rev. B (2011), 83, 184205.